\documentclass[aps,prl,twocolumn]{revtex4}
\usepackage{graphicx}
\usepackage{bm}

\begin{document}

\title{Research in Econophysics}

\author{Victor M. Yakovenko}
\affiliation{Department of Physics, University of Maryland, College
  Park, Maryland 20742-4111}

\date{13 February 2003, {\bf cond-mat/0302270}}

\begin{abstract}
  This article is written for the online newspaper ``The Photon''
  published by the Department of Physics, University of Maryland.  The
  article describes econophysics research done in the group of Victor
  Yakovenko.
\end{abstract} 

\maketitle

\section{Introduction}

During the last several years, in addition to his research in
condensed matter theory, Victor Yakovenko has been working in a
recently emerged field of studies often called ``econophysics''.
Econophysics applies statistical physics methods to economical,
financial, and social problems.  Detailed references to the
econophysics research in Victor Yakovenko's group are given in his Web
page \url{http://www2.physics.umd.edu/~yakovenk/econophysics.html}.
The results have been published in refereed journals
\cite{DY-money,DY-income,DY-wealth,DY-Heston,SY-Heston,DY-statistical}
and presented at international conferences and seminars.  His research
has been reviewed in a popular article in the \textit{American
  Scientist} magazine \cite{Hayes}, and \textit{Australian Financial
  Review}, the leading Australian business newspaper, has published an
op-ed column about his studies \cite{Price}.  His student Adrian
Dragulescu received Ph.D.\ in 2002 and now works as a risk analyst at
the Constellation Energy Group in Baltimore, which is the owner of
Baltimore Electric and Gas Company.  Currently Victor Yakovenko works
with another graduate student A.\ Christian Silva.

\section{Statistical Mechanics of Money, Income, and Wealth}

In this Section, we overview Refs.\ 
\cite{DY-money,DY-income,DY-wealth,DY-statistical}, which use an
analogy with statistical physics to describe probability distributions
of money, income, and wealth.

The equilibrium statistical mechanics is based on the Boltzmann-Gibbs
law, which states that the probability distribution function (PDF) of
energy $E$ is $P(E)=Ce^{-E/T}$, where $T$ is the temperature, and $C$
is a normalizing constant.  The main ingredient in the textbook
derivation of the Boltzmann-Gibbs law is conservation of energy.  When
two economic agents make a transaction, some amount of money is
transferred from one agent to another, but the sum of their money
before and after transaction is the same: $m_1+m_2=m_1'+m_2'$.  Then,
by analogy with statistical physics, the equilibrium PDF of money $m$
in a closed system of agents should have the Boltzmann-Gibbs form
$P(m)=Ce^{-m/T}$, where $T$ is the effective ``money temperature''
equal to the average amount of money per agent.  This exponential
distribution is indeed observed in computer simulations
\cite{DY-money}, as shown in Fig. 1.

\begin{figure}[b]
\includegraphics[width=\linewidth]{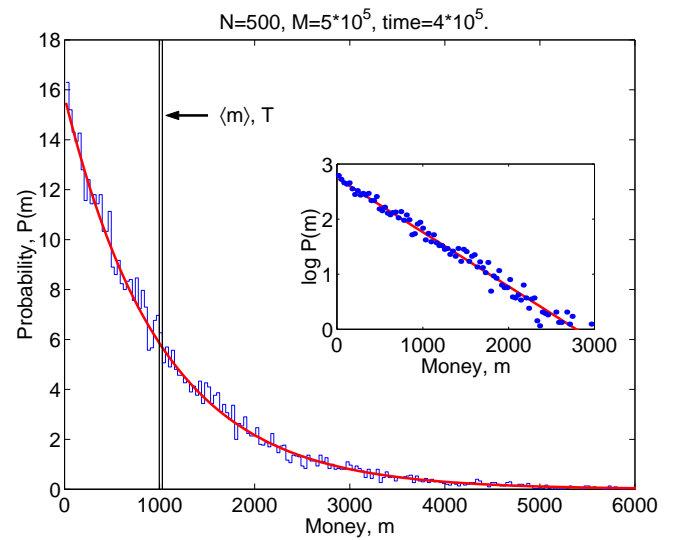} 
\caption{Probability distribution of money in computer 
  simulation \cite{DY-money}.}
\end{figure}

It is interesting to compare this result with the actual PDF of money
in the society.  Unfortunately, it is very difficult to find the data
on distribution of money $m$.  On the other hand, a lot of statistical
data is available for distribution of income $r$ (for revenue).  Fig.
2 shows that the PDF of individual income in USA is very well fitted
by the exponential function $P(r)=Ce^{-r/T}$ \cite{DY-income}.

\begin{figure}[t]
\includegraphics[width=\linewidth]{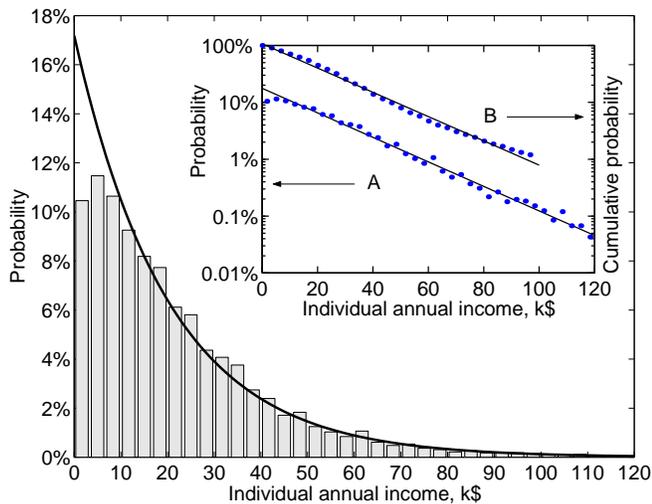} 
\caption{Probability distribution of individual income in USA in 1996 
\cite{DY-income}.}
\end{figure}

The standard plot of PDF inevitably puts an upper limit on the
horizontal axis (120 k\$/year in Fig.\ 2).  A standard way of
representing the whole income distribution without any truncation is
the so-called Lorenz curve shown in Fig.\ 3.  The horizontal axis of
the Lorenz curve, $x(r)$, represents the fraction of population with
incomes below $r$, and the vertical axis $y(r)$ represents the
fraction of the total income this population accounts for.  As $r$
changes from 0 to $\infty$, $x(r)$ and $y(r)$ change from 0 to 1 and
parametrically define the Lorenz curve in the $(x,y)$ space.  The
diagonal line $y=x$ represents the Lorenz curve in the case where all
population has equal income.  The inequality of the actual income
distribution is characterized by the Gini coefficient $0\leq G\leq 1$,
which is the area between the diagonal and the Lorenz curve,
normalized to the area of the triangle beneath the diagonal.  For the
exponential PDF, the Lorenz curve and the Gini coefficient can be
easily calculated \cite{DY-income}:
\begin{equation}
  y = x + (1-x)\ln(1-x),\qquad     G = 1/2.
\label{eq:Lorenz}
\end{equation}
The solid line in Fig.\ 3 shows the theoretical Lorenz curve given by
Eq.\ (\ref{eq:Lorenz}), and the points show the income data for
1979--1997.  The agreement is quite good, in the first approximation,
given that the curve (\ref{eq:Lorenz}) has no fitting parameters.  The
inset shows that the Gini coefficient is close to the theoretical
value 1/2, although the inequality does increase during the last 20
years.

\begin{figure}[b]
\includegraphics[width=0.8\linewidth]{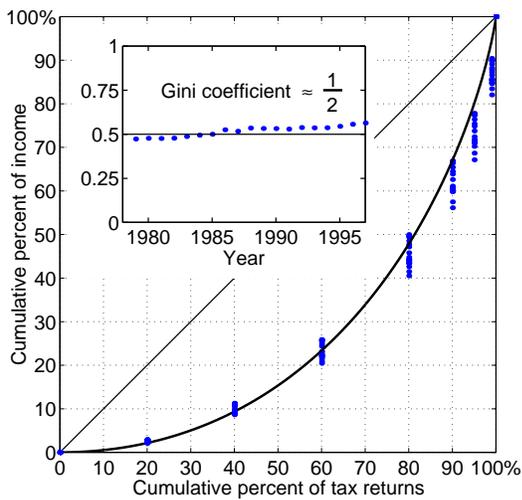} 
\caption{Lorenz curve (main panel) and Gini coefficient (inset).}
\end{figure}

One may notice that discrepancy between the theory and the data occurs
at the upper end of Fig.\ 3.  The origin of this discrepancy becomes
clear when we look at the cumulative distribution of income up to 1
M\$/year shown in Fig.\ 4.

\begin{figure}[t]
\includegraphics[width=\linewidth]{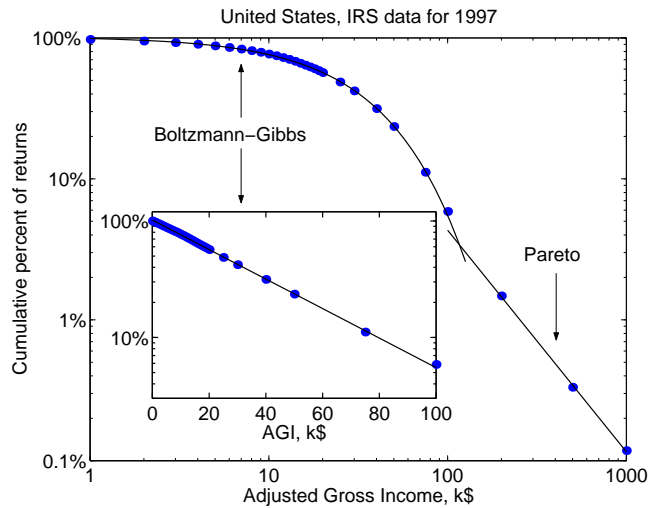} 
\caption{Cumulative probability distribution of individual income 
  in log-log (main panel) and log-linear (inset) scales
  \cite{DY-statistical}.}
\end{figure}

It is clear from Fig.\ 4 that income distribution for the great
majority of population (more than 97\%) is described by the
exponential Boltzmann-Gibbs law.  However, for a small fraction of
population (less than 3\%) with income above 100 k\$/year, the PDF
changes to the Pareto power law.  The extra income in the upper tail
of the distribution can be considered as a ``Bose condensate'', and the
Lorenz curve should be modified as \cite{DY-statistical}
\begin{equation}
  y = (1-f)\,[x + (1-x)\ln(1-x)] + f\,\delta(1-x),
\label{eq:Bose}
\end{equation}
where the last term is the delta-function, and $f$ is the fraction of
income in the ``Bose condensate''.  As shown in Fig.\ 5, Eq.\ (2) gives
an excellent fit of the data, and $f=16\%$ in 1997.

\begin{figure}[b]
\includegraphics[width=0.8\linewidth]{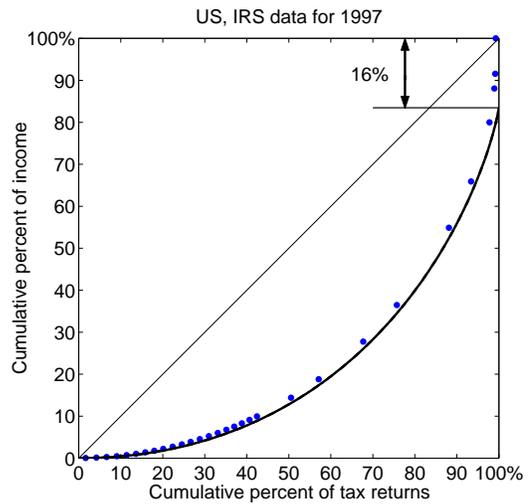} 
\caption{Lorenz curve given by Eq.\ (\ref{eq:Bose}).}
\end{figure}

Thus far we discussed the distribution of individual income.  By
taking a convolution of two exponential distributions, it is easy to
show that the PDF of family income is given by the modified
exponential formula $P(r)=Cre^{-r/T}$ \cite{DY-income}.  As Fig.\ 6
shows, this formula is in excellent agreement with the data.  The
corresponding Lorenz curve for family income is shown in Fig.\ 7 and
compared with the data from the Bureau of Census for 1947--1994.  It
is amazing that the shape of the income distribution remains the same
for half a century, and it is in agreement with the theoretical
formula.

\begin{figure}[t]
\includegraphics[width=0.95\linewidth]{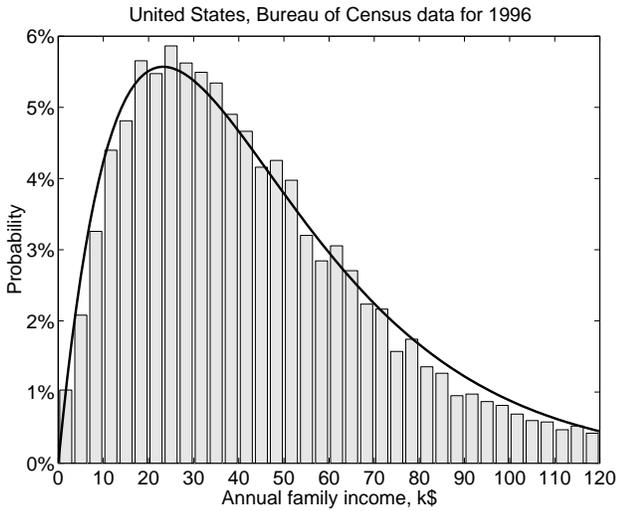} 
\caption{Probability distribution of family income.}
\end{figure}

\begin{figure}[b]
\includegraphics[width=0.8\linewidth]{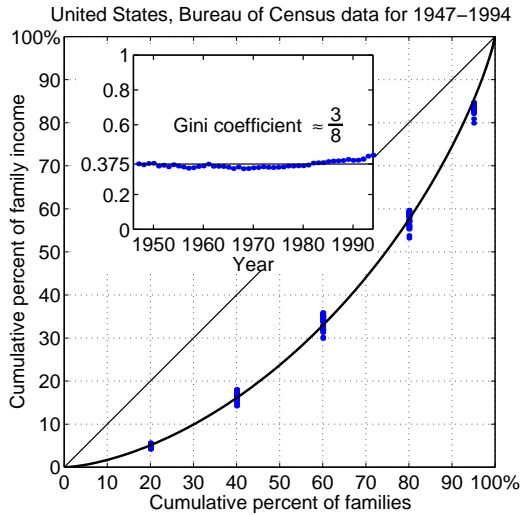} 
\caption{Lorenz curve and Gini coefficient for family income.}
\end{figure}

The theoretically calculated Gini coefficient for family income is
3/8=37.5\% \cite{DY-income}.  The inset in Fig.\ 7 shows that the Gini
coefficient in USA reported by the Bureau of Census for 1947--1994 is
very close to the theoretical value.  The average Gini coefficients
for different regions of the World in 1988 and 1993 are shown in Fig.
8.  For the well-developed market economies of West Europe and North
America, the Gini coefficient is very close to the calculated value
37.5\% and does not change in time.  In other regions of the World,
income inequality is higher.  The special case is the former Soviet
Union and East Europe, where inequality was lower before the fall of
communism and has greatly increased afterwards.

\begin{figure}[t]
\includegraphics[width=\linewidth]{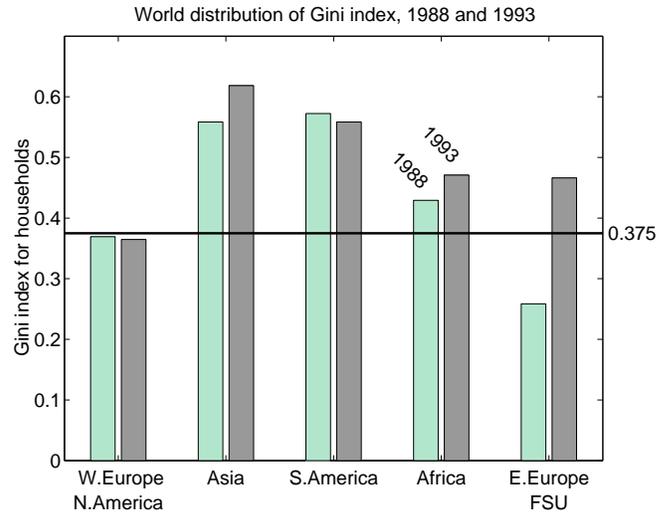} 
\caption{Gini coefficients for different regions of 
  the World \cite{Milanovic}.}
\end{figure}

In statistical physics, the exponential Boltzmann-Gibbs distribution
is the equilibrium one, because it maximizes entropy.  The data shown
above demonstrate that probability distribution of income is also
described by the Boltzmann-Gibbs law, and the equilibrium state of
maximal entropy has been achieved in developed market economies.

\section{Probability Distribution of Stock-Market Fluctuations}

The first theory of stock-market fluctuations was proposed in 1900 in
the Ph.D. thesis of the French mathematical physicist Louis Bachelier
\cite{Taqqu}.  (Henri Poincaré was on his Ph.D.\ committee.)  His
thesis developed the concept of Brownian motion (before the famous
Einstein's paper of 1905) for stock-market prices.  A modern version
of this theory is routinely used in financial literature.  The theory
predicts a Gaussian probability distribution for stock-price
fluctuations.  On the other hand, it is well known that the tails of
the distribution are not Gaussian (the so-called ``fat tails'').  To
improve agreement, it was proposed that the diffusion coefficient of
the Brownian motion is not a constant, but itself is a stochastic
variable.  One popular model was proposed by Steve Heston, who is a
faculty of the Department of Finance, University of Maryland.

\begin{figure}[t]
\includegraphics[width=\linewidth]{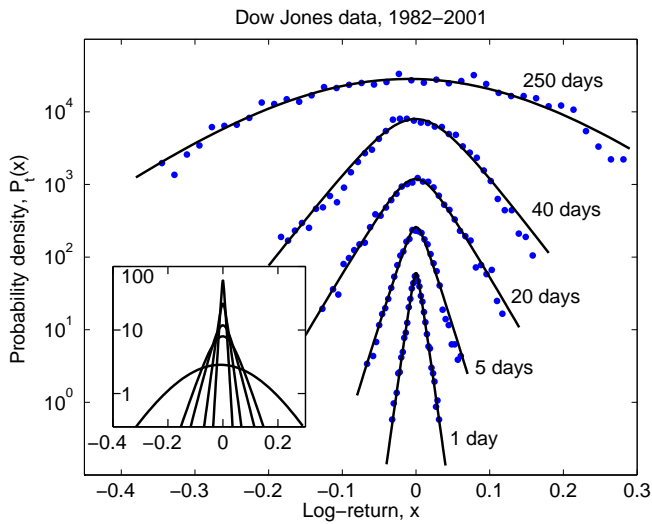} 
\caption{Probability distribution of log-return $x$ for different time 
  lags $t$.}
\end{figure}

\begin{figure}[b]
\includegraphics[width=\linewidth]{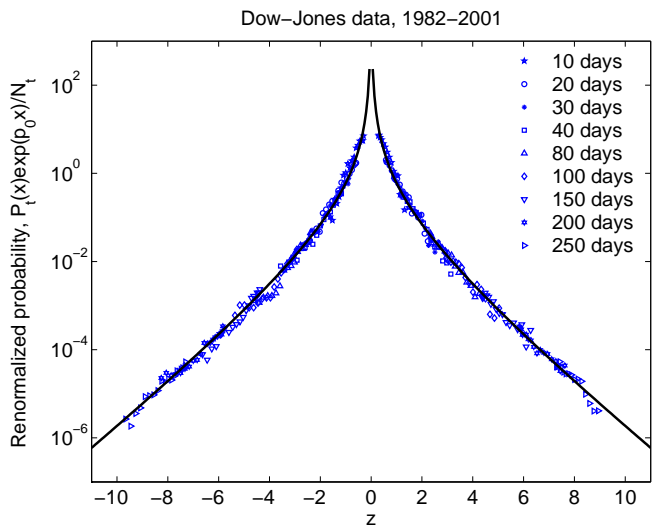} 
\caption{Scaling plot of the probability distribution of log-return $x$.}
\end{figure}

In Ref.\ \cite{DY-Heston}, Dragulescu and Yakovenko derived the
probability distribution of price changes for the Heston model and
compared it with the data.  Fig.\ 9 shows the probability distribution
of log-return $x$ for the Dow-Jones index during the 20-years period
1982--2001.  (In the main panel of Fig.\ 9, the curves are offset
vertically for clarity; the inset shows the same curves without
offset.)  The log-return $x=\ln(S_2/S_1)$ is the logarithm of the
ratio of the stock prices $S_2$ and $S_1$ for two moments of time
$t_2$ and $t_1$ with the average market growth subtracted.  The
probability distribution $P_t(x)$ depends on the time lag $t=t_2-t_1$,
which is indicated near each curve in Fig.\ 9.  The solid curves show
the analytically derived distribution, and the points show the
Dow-Jones data.  We see that the Dragulescu-Yakovenko formula
\cite{DY-Heston} very well describes $P_t(x)$ for a broad range of
time lags from one day to one year (252 trading days).

Dragulescu and Yakovenko also found that for times $t$ longer than the
relaxation time of the model, $P_t(x)$ becomes a function of a single
combination $z$ of the two variables $x$ and $t$.  Thus, when plotted
vs.\ $z$, the points for different time lags should collapse on a
single scaling curve.  That indeed happens, as show in Fig. 10.  The
solid line is the theoretically calculated scaling curve expressed in
terms of a modified Bessel function.  Notice that the agreement
extends over \textit{seven} orders of magnitude on the vertical axis.

In the recent paper \cite{SY-Heston}, Silva and Yakovenko found that
the same results hold for Nasdaq and S\&P 500 in 1980s and 1990s.  By
analyzing the statistics of fluctuations, they concluded that the
decline of stock market after 2000 is a long-term change of regime,
not a temporary fluctuation, unlike the crash of 1987.

\section{Conclusions}

We have demonstrated that methods and techniques of statistical
physics can be successfully applied to economical and financial
problems.  The great experience of physicists in working with
experimental data gives them a unique advantage to uncover
quantitative laws in the statistical data available in economics and
finance.  The interdisciplinary field of econophysics is bringing new
insights and new perspectives, which are likely to revolutionize the
old social disciplines.

\end{document}